\documentclass[12pt,a4paper]{article}
\usepackage{graphicx}

\newcommand{\disable}[1]{}

\newcommand{\Lower}[2]{\smash{\lower #1 \hbox{#2}}}
\newcommand{\alice}{\textsl{ALiCE}~}
\newcommand{\be}{\begin{equation}}
\newcommand{\ee}{\end{equation}}
\newcommand{\ba}{\begin{eqnarray}}
\newcommand{\ea}{\end{eqnarray}}
\newfont{\bb}{msbm10}
\def\Real{\mbox{\bb\symbol{82}}}
\def\Complex{\mbox{\bb\symbol{67}}}

\newcommand{\R}{\Real}
\newcommand{\C}{\Complex}

\newcommand{\umx}{U_{\mu\,x}}

\newcommand{\xpm}{x+\hat\mu}
\newcommand{\xmm}{x-\hat\mu}

\newcommand{\umxmm}{U_{\mu\,\xmm}}

\newcommand{\journal}[5]{#1;\textsl{ #2} \textbf{#3 }(#4) #5}

\newcommand{\eqref}[1]{(\ref{#1})}

\newcommand{\gral}{\textsl{GRAL}~}

\title{On the scaling of computational particle physics codes on cluster computers} 
\author{Z.~Sroczynski\\
\textsl{\small Division of Theoretical Physics, Dept.~of Mathematical
  Sciences}\\
\Lower{-1ex}{\textsl{\small University of Liverpool, Liverpool L69 3BX, UK}}\\
N.~Eicker, Th.~Lippert, B.~Orth and K.~Schilling\\
\textsl{\small Theoretical Physics, Wuppertal University}\\
\Lower{-1ex}{\textsl{\small Gau{\ss}stra{\ss}e 20, D-42097 Wuppertal, Germany}}}

\begin{document}
\date{}
\maketitle

\begin{abstract}
  Many appplications in computational science are sufficiently
  compute-intensive that they depend on the power of
  parallel computing for viability. For all but the ``embarrassingly parallel''
  problems, the performance depends upon the level of granularity
  that can be achieved on the computer platform.    

  Our computational particle physics
    applications require machines that can
  support a wide range of granularities, but in general,
  compute-intensive state-of-the-art projects will require 
  finely grained distributions.
  Of the different types of
  machines available for the task, we consider cluster computers.

  The use of clusters of commodity computers in high  performance
  computing has many advantages including the raw  price/performance
  ratio and the flexibility of machine configuration and  upgrade. 
  Here we focus on what is usually considered the weak point of cluster
  technology; the scaling behaviour when faced with a numerically
  intensive parallel computation. 
  To this end we examine the scaling of our own applications from
  numerical quantum field theory on a
  cluster and infer conclusions about the more general case.
\end{abstract}


\section{Introduction}\label{intro}

A compute-intensive application might have to be distributed over many
compute nodes in order for it to run at a viable speed. One  needs to
know how well a platform can support the required level of granularity.
In this paper we examine this in the case of cluster computers.

The use of clusters of computers in high performance computing has
achieved some success with the price/performance 
advantage of using low-cost and yet high-speed commercial
off-the-shelf (COTS) processors to build a fast machine with the
budget available. The significant presence of clusters in the TOP500
list \cite{top500} attests to the power of this approach.

Cluster computers are also becoming more widely accepted with the
growing ease with which 
they
 can be configured and maintained; there are now numerous open source
and commercial ``Cluster in a box'' packages available. It can also be
important that portable application code can be developed and run on these
machines, using standard operating systems, libraries, tools and parallel
APIs.

Essential for the use of clusters in many areas of high performance computing
is the presence of high performance system area network technology. 
The question is whether this is good enough to make parallel
computation on clusters viable in the face of very fast
communications built into MPP or SMP/NUMA machines. 

In this paper we intend to make some quantitative statements on the
subject by examining the scaling of some codes from our parallel application,
\textit{viz.} Lattice QCD. In section \ref{lqcd} we give a brief
description of Lattice QCD, emphasising the computational task involved.
Then in section \ref{lqcdsoft} we describe how our code implements
this with particular reference to the parallelisation. 
We proceed to describe in section \ref{tests} the scaling tests we
apply and present their results.
These are  summarised and discussed in section \ref{summary}, after
which we present our conclusions.

\section{Lattice QCD}\label{lqcd}

Lattice QCD is a numerical evaluation of Quantum Chromodynamics, 
the theory of the strong interaction, which takes its name from the
use of a four-dimensional hypercuboid lattice to represent a
discretised spacetime (see \textit{e.g.} \cite{rajan, suss} for a recent
overview of the subject). In the lattice theory, each lattice site
(vertex of the lattice) represents a spacetime point.

These computations are demanding enough that they routinely make much
use of supercomputers, and the grid-based  formulation of the problem
makes it most suitable for data decomposition on parallel machines.

One of the central defining features of a Lattice QCD calculation is
the discretised representation of the Dirac fermion operator which
describes the interaction of quarks in the theory. On the lattice,
the discretised operator takes the form of a fermion \textsl{matrix}. 
The algorithms used for Lattice QCD calculations are such that the
bulk of the computer time is spent in the
inversion of this matrix.
There 
is a variety of representations of the fermion matrix which
have different properties on the lattice but they should all 
become identical
to the continuum Dirac operator in the limit of zero lattice spacing.
The particular formalism used in this work is that originally due to Wilson
\cite{wilson}. The Wilson fermion matrix is 
\be
M_{xy} =  \delta_{xy}-\kappa D_{xy}
\ee
where the indices $x$ and $y$ refer to lattice sites and
$\kappa\in\R$ is the \textsl{hopping parameter}, related to
the bare quark mass.

The Wilson \textsl{hopping matrix} is
\be
D_{xy} = \sum_{\mu=0}^3(1-\gamma_{\mu})\umx\delta_{y\,\xpm}
+(1+\gamma_{\mu})\umxmm^{\dagger}\delta_{y\,\xmm}
\ee
where
the lattice \textsl{gauge fields} $\umx\in SU(3) \subset\C^{3\times 3}$
carry \textsl{colour} indices and
the Dirac $\gamma$-matrices $\gamma_\mu\in\C^{4\times 4}$:
$\{\gamma_\mu,\gamma_\nu\} = 2\delta_{\mu\nu}$ carry \textsl{spin} indices.

Therefore $M$ is a matrix in lattice site $\otimes$ spin $\otimes$
colour space. If the number of sites in the lattice is $V$ then the
fermion matrix is a complex matrix of size $(12V)^2$ acting on vectors
consisting of $12V$ complex numbers.

The lattice is of finite extent, and we impose periodic or antiperiodic
boundary conditions.

Note that the lattice site structure of this matrix is either local or
connects nearest-neighbour sites.
Therefore it is a sparse matrix and is
implemented as an operator rather than stored explicitly

Other formulations of the fermion matrix include the
\textsl{Sheikholeslami-Wohlert} 
discretisation which adds to the Wilson matrix a $3\otimes4\times3\otimes4$
complex matrix diagonal in lattice sites \cite{clover} , variations of the
\textsl{Kogut-Susskind} matrix \cite{staggered} which have no spin components
but can have a more than nearest-neigh\-bour connectivity \cite{morestaggered,
  yetmorestaggered} and the \textsl{Fixed Point} formulations which have spin
and colour indices and more than nearest-neighbour connectivity
\cite{perfect}. The choice depends on the desired properties, and has
consequences for the balance of computation to communication required; for
example, the Sheikoleslami-Wohlert matrix has more local computation than the
Wilson matrix with no extra communication and will therefore scale better.

The Wilson matrix has the favourable property of only requiring
nearest-neigh\-bour communications but since its computational requirements are
comparatively modest it does put some pressure on the scaling abilities of a
parallel machine. Although it is one of the older formulations it is still
interesting not least because it 
can be used as the basis for the new breed of \textsl{Ginsparg-Wilson} lattice
fermion formulations \cite{gw, overlap}, which have very
favourable field-theoretic properties but are still very
expensive to compute with; a single multiplication by the GW fermion matrix
involves many ($\mathcal{O}(100)$) multiplications by the Wilson matrix.  For
these reasons we believe that our test results with the Wilson matrix are
relevant. However we also consider the effects of a fermion matrix requiring
more than nearest-neighbour communications.



\section{Description of Lattice QCD software}\label{lqcdsoft}


We give here some details of the implementation of our lattice QCD 
applications, describing the computational task itself, how the code
is written and how the code is parallelised.

Much of the code used in these tests was actually developed for
a specific Lattice QCD project called \gral
\footnote{\gral\ stands for ``Going
  Realistic And Light''.} 
\cite{gral}.
The \gral project has as its aim the finite size scaling analysis of QCD
physics close to the chiral regime of light dynamical quarks 
using the Wilson formulation.  This requires a large computational effort, and
although a cluster is just one of the machines that is used for the project,
good performance of the Lattice QCD code on it is important to 
the project's success.

The bulk of the code is written in C compiled with the Compaq C
Compiler \texttt{ccc} with a set of optimisations enabled with the
\texttt{-fast} flag. 
We use assembly language kernels \cite{akmt} for certain low-level numerically
intensive routines where performance is critical. 
Routines from the optimised BLAS library are also used where appropriate.
Both 32-bit and 64-bit code is implemented, but since 64-bit accuracy
is preferred for the \gral project, we present here only 64-bit results. 

The communications are entirely handled with message passing \textit{via} MPI.
The fast MPI-library on ALiCE is provided by the
ParaStation\texttrademark\ 
cluster middleware \cite{parastation}.

\subsection{Fermion matrix inversion}\label{solvers}

As stated in section \ref{lqcd} most computer time 
will be spent in those portions of this code which invert the fermion
matrix. This involves solving the linear system
\be
M\psi = b
\ee
where $M$ is the fermion matrix defined in section \ref{lqcd}, $b$ is
the source vector and  $\psi$ is the solution to be found.
A Krylov subspace algorithm is best suited to the solution of such a
sparse linear system.
We have implemented several such algorithms but here we show results using the
BiCGStab \cite{bicgstab, morebicgstab} method which is usually the fastest. The
communication requirements (described in section
\ref{nn-commreq}) of these different algorithms are comparable.

An additional refinement is the use of a parallel (locally
lexicographic) SSOR preconditioning
\cite{llssor, morellssor} to reduce the number of iterations of the BiCGStab
solver. This requires a slightly different pattern of communication
and so we present these results separately.

\subsection{Grid decomposition}

The four-dimensional lattice is decomposed over a grid of processing
nodes. This grid can be from zero dimensions 
(\textit{i.e.,} a single node) to
four dimensions (since the lattice is four-dimensional). 
The number of nodes along any dimension is restricted 
in that the local lattice on each node must have identical size and
shape. This constraint is not strictly necessary on a
MIMD machine but it makes the implementation simpler and ensures 
perfect load balancing.

\subsection{Nearest-neighbour communication requirements}\label{nn-commreq}

Every iteration of the solver algorithm requires 
three global summations of a complex number and
performs two multiplications by the fermion matrix.

\begin{figure}
\begin{center}
    \includegraphics[width=.83\textwidth]{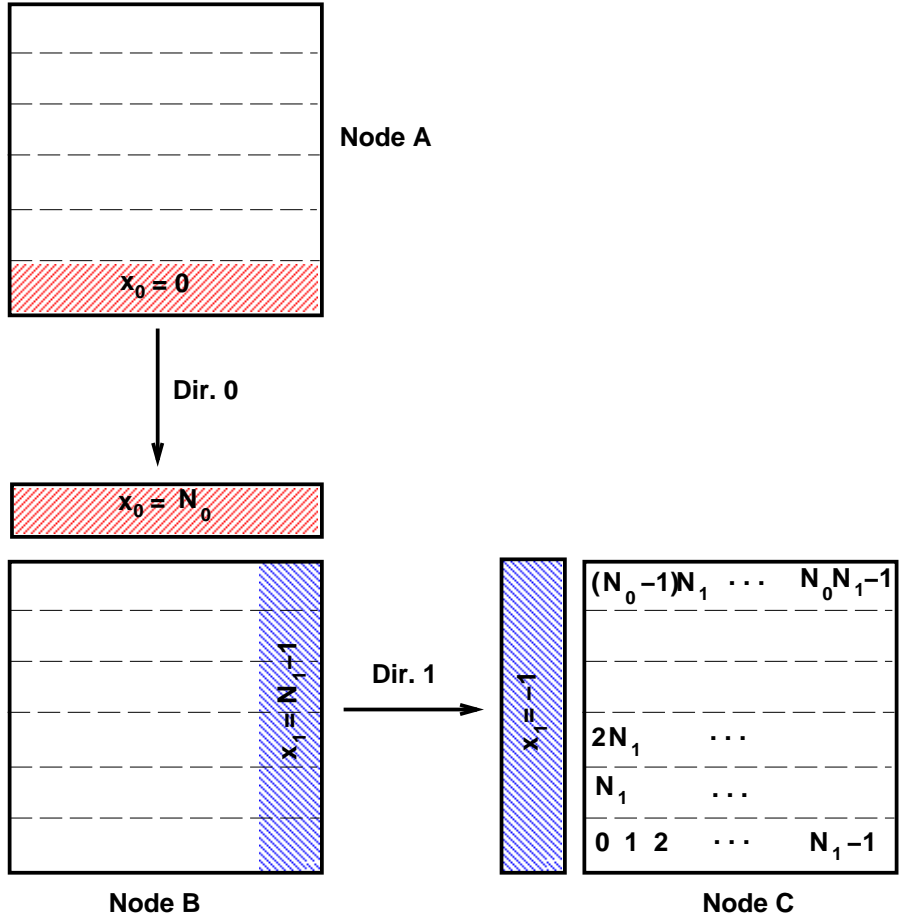}
  \caption{\label{nn-comms}An illustration of the nearest-neighbour
    communications in two dimensions of a two-dimensional local lattice of
    size $N_0\times N_1$.  Nodes A and B show the communication in a
    1-dimensional processor grid in the negative direction 0: The data on Node
    A with coordinate $x_0=0$ is copied to a buffer on Node B where it can be
    accessed as Node B data with coordinate $x_0=N_0$. The positive direction
    0 communication is analogous.  Nodes C and B show the communication in a
    2-dimensional grid in the positive direction 1: The data on Node B with
    coordinate $x_1=N_1-1$ is copied to a buffer on Node C where it can be
    accesssd as local data with coordinate $x_1=-1$. The negative direction 1
    communication is analogous.  Each node sends data in this way from all its
    boundaries.  }
\end{center}
\end{figure}

Each fermion matrix multiplication requires nearest-neighbour
communication of the multiplicand vector. 
On a lattice boundary in a non-parallelised direction this is taken care
of by the boundary condition. In a parallelised direction the
required data is on another node.
With the message passing communication paradigm, 
this means that the data on those lattice sites on the boundary of 
a local lattice is sent to the neighbouring node. Each node receives
data from neighbouring nodes and stores it in buffers.
This communication pattern is illustrated in figure \ref{nn-comms}.

The amount of data to be sent depends on the size of the lattice and
the grid decomposition. If these are such that
 on each node of a grid of dimensionality $d>0$
the local lattice has $V$ sites,
then the number of sites on the boundaries  from which data must be
sent is
\be \label{commsurface}
2V\sum_{\mu=0}^{d-1}\frac{1}{N_\mu},
\ee
where the sum is over parallelised directions $\mu$ and $N_\mu$ is the length
of the local lattice in direction $\mu$.  Recall that a vector consists of 12
complex numbers on each lattice site, which is 192 bytes at 64-bit precision.

For the normal BiCGStab algorithm  all the data on a boundary can be
sent at once as a single message. We send each boundary 
before the multiplication by the fermion matrix begins. The order in
which they are sent is essentially unimportant. 

The SSOR preconditioning requires most of the data to be sent incrementally as
a succession of smaller messages at various points within the
matrix-vector
multiplication; the order is important.  This arrangement is potentially less
favourable because of a possibly inefficient use of available bandwidth and
the effect of multiple latency penalties.

\subsubsection{Data layout}\label{nn-layout}


The parallelisation efficiency is affected by the way the data is stored along
with the design of the communication buffers.

The vectors involved in the calculation are stored as site-major arrays. Each
array element contains the 12 complex numbers and the 
$n-$th 
element
corresponds to the local lattice site with cartesian coordinates $(x_0, x_1,
x_2, x_3)$ \textit{via} $n = x_3+N_3x_2+N_3N_2x_1+N_3N_2N_1x_0$ where $N_\mu$
is the size of the local lattice in the $\mu$th direction.  The gauge fields
are stored in a similar way.

This storage order is 
illustrated in figure \ref{nn-comms}. 
Note that because of the layout of the data in lattice, the data on
the direction 1 
boundaries is strided, which takes more time to copy than
contiguous data. 

The communication buffers are, for convenience, attached to this array,
which increases its length by an amount the size of the local lattice
boundaries. Two arrangements of the buffers were tried:

\textbf{The ``halo'' layout}

The ``halo'' layout places the buffers on either
side of the array body, \textit{e.g.} for a local lattice of size 
$N_0\times N_1\times N_2\times N_3$, the first two buffers would be

\begin{center}
\fbox{\strut $x_1=-1$}\fbox{\strut $x_0=-1$}\fbox{\strut $n=0\cdots N_0N_1N_2N_3-1$}\fbox{\strut $x_0=N_0$}\fbox{\strut $x_1=N_1$}
\end{center}

\textbf{The ``tail'' layout}

An alternative is the ``tail'' layout: 

\begin{center}
\fbox{\strut $n=0\cdots N_0N_1N_2N_3-1$}\fbox{\strut $x_0=-1$}\fbox{\strut $x_0=N_0$}\fbox{\strut $x_1=-1$}\fbox{\strut $x_1=N_1$}
\end{center}

where the buffers are attached to the end of the array. In section \ref{nn-layout-results} we
test the differences between these two layouts.

\subsection{Next-to-nearest-neighbour communication\\ requirements}\label{nnn-commreq}  

Although the Wilson fermion matrix requires only nearest-neighbour
communication of the gauge field $U$, for algorithmic reasons we also need
next-to-nearest-neighbour access%
\footnote{
 In a typical \gral application the
gauge field is communicated $\mathcal{O}(100)$ times less often than the
spin-colour vector, therefore the efficiency of the gauge field communications is not so
important for our considerations.}.

\begin{figure}
\begin{center}
    \includegraphics[width=.83\textwidth]{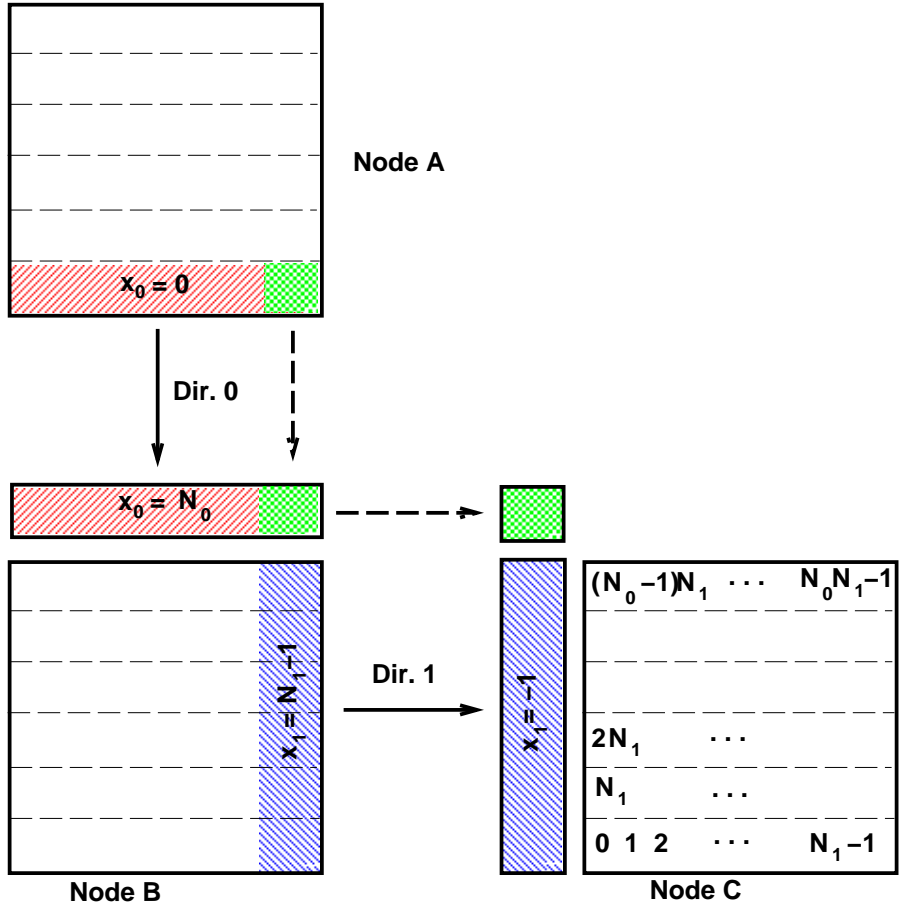}
  \caption{\label{nnn-comms}An illustration of next-to-nearest-neighbour
  communications in two dimensions of a two-dimensional local lattice of size
  $N_0\times N_1$. 
In addition to the data communications shown in
  figure \ref{nn-comms} 
the corner of the node A lattice (coordinates $x_0=N_0$, $x_1=N_1-1$)
is sent to a buffer on node C 
  where it can be accessed as data with coordinates $x_1=-1$, $x_0=-1$.
We do this in two stages, sending the data \textit{via} the $x_0=N_0$
  buffer of node B, so the direction 0 communication must complete
  before the direction 1 
communication can start. Blocking or semi-blocking semantics are suitable here.
An alternative (not examined here) would be to send the corner
 from node A directly to node C, which in principle avoids this wait.
}
\end{center}
\end{figure}

Furthermore, as mentioned in section \ref{intro}, some fermion matrices
connect more than nearest-neighbour lattice sites. We investigate the effects
of this by considering next-to-nearest-neighbour communication.  This is
already possible on a 1-dimensional grid with the communication patterns
already presented, but on a 2-dimsional grid the communication in direction 1
has to be changed as shown in figure \ref{nnn-comms}. Additional data from the
corners of the lattices has to be sent.

For a lattice with $V$ sites on each node of a grid of
dimensionality $d>1$, there are 
\be
4V\sum_{\mu=0}^{d-1}\sum_{\nu=0}^{\mu-1}\frac{1}{N_\mu N_\nu}
\ee
lattice sites on the corners  from which data must be sent, where
the sum is over parallelised directions $\mu$ and $\nu$ and
 $N_\mu$ and $N_\nu$ are the lengths of the local lattice in those directions. 
This is in addition to the amount of boundary data described in
section \ref{nn-commreq}.
We wish to test the effect of this additional communication. 

\subsubsection{Data layout}\label{nnn-layout}

The next-to-nearest neighbour communication pattern in more than one
dimension requires extra buffers on each array for the corner data.  The
possibilities for different data layouts are therefore increased. We
illustrate three possibilities for the layout on a two-dimensional
grid (which can be generalised for a greater number of dimensions):

\textbf{Layout 0}

The corner buffers are added as halos to the $x_1=-1$ and
$x_1=N_1$ buffers;

\fbox{\strut $x_1=-1, x_0=-1$}\fbox{\strut $x_1=-1$}\fbox{\strut $x_1=-1, x_0=N_1$}

\parbox[b]{\textwidth}{\centering\fbox{\strut $x_0=-1$}\fbox{\strut $n=0\cdots N_0N_1N_2N_3-1$}\fbox{\strut $x_0=N_0$}}

\hfill\fbox{\strut $x_1=N_1, x_0=-1$}\fbox{\strut $x_1=N_1$}\fbox{\strut $x_1=N_1, x_0=N_1$}
  
\textbf{layout 1}

The corner buffers are added to the $x_1=-1$ and
$x_1=N_1$ buffers as tails;

\fbox{\strut $n=0\cdots N_0N_1N_2N_3-1$}\fbox{\strut $x_0=-1$}\fbox{\strut $x_0=N_0$}

\parbox[b]{\textwidth}{\centering\fbox{\strut $x_1=-1$}\fbox{\strut $x_1=-1, x_0=-1$}\fbox{\strut $x_1=-1, x_0=N_1$}}

\hfill{\fbox{\strut $x_1=N_1$}\fbox{\strut $x_1=N_1, x_0=-1$}\fbox{\strut $x_1=N_1, x_0=N_1$}}

\textbf{layout 2}

The corner buffers are added as a tail at the end of the array;

\parbox[b]{\textwidth}{
\fbox{\strut $n=0\cdots N_0N_1N_2N_3-1$}\fbox{\strut
  $x_0=-1$}\fbox{\strut $x_0=N_0$}\fbox{\strut $x_1=-1$}\fbox{\strut
  $x_1=N_1$}\\

\hfill\fbox{\strut $x_1=-1,x_0=-1$}\fbox{\strut $x_1=-1, x_0=N_1$}\fbox{\strut $x_1=N_1, x_0=N_1$}\fbox{\strut $x_1=N_1, x_0=-1$}
}

Layouts 0  and  1 have the advantage that the data in the $x_0=N_0$
and $x_0=-1$ buffers with $x_1=0\;(N_1-1)$
  can be sent in the
negative (positive) direction 1 along with the  data with
$0\leq x_0<N_0\mbox{, }x_1=0\;(N_1-1)$ as a single message.

Layout 2 requires that the corners buffers be sent seperately. 
However, since it
is merely the tail layout of section \ref{nn-layout} with extra buffer
space added to the end of it, 
a single layout can be used for fields both with and without
next-to-nearest-neighbour interaction without wasting space.



\section{The scaling tests}\label{tests}

Our goal is to test how well the code introduced in section \ref{lqcdsoft}
scales on a cluster. We run tests to see how this is affected by various
factors; the data layout, the locality of the fermion matrix and the solver
algorithm.

In this section we first describe the essential details of all the scaling
tests.  We then describe the hardware and software on which these tests are
run.

\subsection{The test specifications}\label{testspec}

In order to test the scaling of our lattice QCD applications, we
choose to test the fermion matrix inversion described in section \ref{solvers}
since it is the area in which many QCD programs and in particular the \gral
applications will spend most run-time and it has the considerable
communication requirements discussed in section \ref{nn-commreq}.  All the
tests are run in 64-bit precision.

We choose to time the inversion on a $12^4$ lattice. This rather arbitrary
size is smaller than that used in most realistic lattice QCD calculations.
As our machine size is 128 processors, this lattice size
should give an idea of how the performance is affected by the size and
surface/volume ratio of the sub-lattices on each node
when extrapolated to larger computers with 512 or 1024 processors.

Each solve is performed on a random $SU(3)$ gauge field with a random gaussian
source vector. The solve is repeated many times until a good estimate of the
average run-time per solve can be obtained.  The run-time on a single node is
very well reproducible, but on multiple nodes there is some variation, perhaps
due to effects of different physical grid topologies or the load on the
network and switches, so the entire run is repeated at least five times to
average over such variations.

Table \ref{partab} shows the grid topologies used in the tests. 
Results from grids of the same dimensionality and number of nodes are averaged.
The ratio of the communicable surface of the local lattice to the
local lattice volume is also shown.

\begin{table}\centering
\begin{tabular}{cccc}
\hline 
nodes & grid & local lattice & surface/volume\\
\hline
2 & $2\times1\times1\times1 $&$ 6\times12\times12\times12 $& 0.33 \\
3 & $3\times1\times1\times1 $&$ 4\times12\times12\times12 $& 0.50 \\
4 & $4\times1\times1\times1 $&$ 3\times12\times12\times12 $& 0.67 \\
6 & $6\times1\times1\times1 $&$ 2\times12\times12\times12 $& 1.00 \\
12 & $12\times1\times1\times1$&$  1\times12\times12\times12$&  2.00\\
\hline
4 & $2\times2\times1\times1 $&$ 6\times6\times12\times12  $&0.67  \\
9 & $3\times3\times1\times1 $&$ 4\times4\times12\times12  $&1.00  \\
12 & $4\times3\times1\times1$& $3\times4\times12\times12 $&  1.17 \\
12 & $3\times4\times1\times1$& $4\times3\times12\times12 $&  1.17 \\
16 & $4\times4\times1\times1 $&$ 3\times3\times12\times12  $&1.33  \\
36 & $6\times6\times1\times1 $&$ 2\times2\times12\times12  $&2.00  \\
\hline
 8 & $2\times2\times2\times1$ & $6\times6\times6\times12$ & 1.00  \\
27 & $3\times3\times3\times1$ & $4\times4\times4\times12$ & 1.50  \\
36 & $3\times3\times4\times1$ & $4\times4\times3\times12$ & 1.67  \\
64 & $4\times4\times4\times1$ & $3\times3\times3\times12$ & 2.00  \\
\hline
16 & $2\times2\times2\times2$ & $6\times6\times6\times6$ & 1.33 \\
36 & $2\times3\times3\times2$ & $6\times4\times4\times6$ & 1.67 \\
64 & $2\times4\times4\times2$ & $6\times3\times3\times6$ & 2.00 \\
\hline
\end{tabular}
\caption{\label{partab}The grid sizes, local lattice sizes and local
  lattice surface/volume ratios used in the scaling measurements. The global
  lattice size is $12^4$ throughout our tests.}
\end{table}

Our main metric of  the parallel performance is the \textsl{speedup};
\be
\mbox{speedup} = \frac{\mbox{speed on \textit{n} nodes}}{\mbox{speed on 1 node}}.
\ee
It can also be useful to know the 
\textsl{parallel efficiency} (\textsl{scaled speedup});
\be
\mbox{parallel efficiency} 
=
\mbox{scaled speedup} 
= \frac{\mbox{speedup}}{\mbox{\textit{n}}}.
\ee

\subsection{The test platform}\label{alice}

\textsl{ALiCE}, the Alpha-Linux Cluster Engine at Wuppertal University is a
general purpose high-performance computer consisting of 128 Compaq
DS10 servers, each with a 616MHz Alpha 21264 processor,  64 kB 
2-way set-associative 
level 1 data and instruction caches and a 2 MB level 2 cache \cite{alice}. 
The Alpha 21264 is 4-way superscalar, sustaining 2 floating-point
operations per cycle. The operating system is SuSE
Linux \cite{suse}.

\alice is clustered using Myrinet with the ParaStation3 cluster
middleware\cite{parastation}.  This arrangement takes advantage of the open
architecture of the Myrinet communication hardware to replace the standard
firmware supplied by Myricom with that supplied by Par-Tec, together with
kernel extensions and communications libraries.  The ParaStation system
endeavors to achieve high performance by implementing a reliable
communications protocol in firmware and keeping communications in user space.
The two-way 1.28 Gbit/s Myrinet network is configured as a multistage crossbar
and the Myrinet cards have a 64 bit/33 MHz PCI buses.  We use MPICH 1.2.3 as
the interface between our application code and the ParaStation middleware.  On
\textsl{ALiCE}, the latency for ParaStation/MPI is $16 \mu$s; the bi-directional
bandwidth is about 170 MB/s \cite{parallelfilesystem}.



\subsection{Single node performance}\label{singlenode}

\begin{figure}
  \begin{center} 
    \includegraphics[width=.83\textwidth]{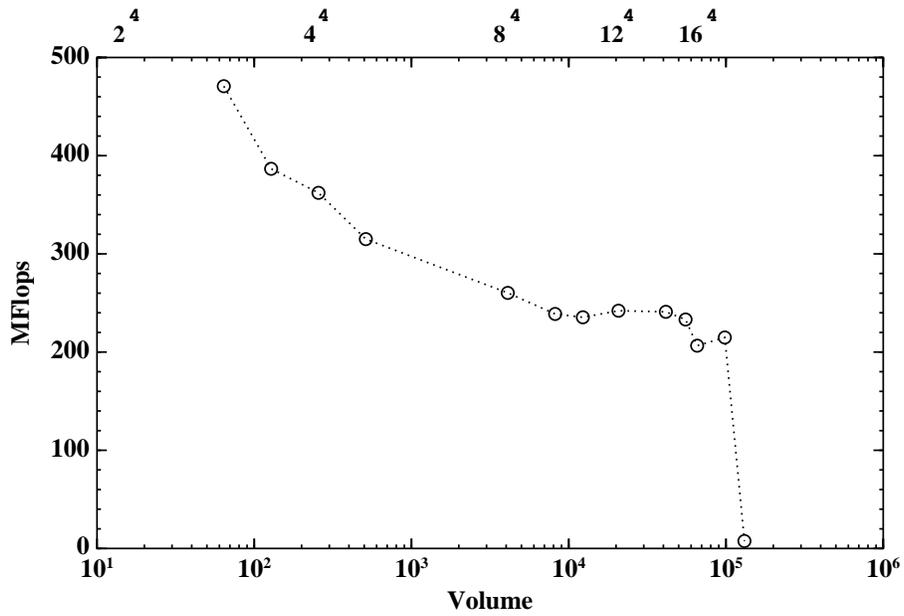}
    \caption{\label{bicgstab-singlenode}
      The speed of the BiCGStab solver on a single node as a function
      of the lattice volume. 
      The corresponding lattice dimensions are indicated along the top.
    }
  \end{center} 
\end{figure}

\begin{figure}
  \begin{center} 
    \includegraphics[width=.83\textwidth]{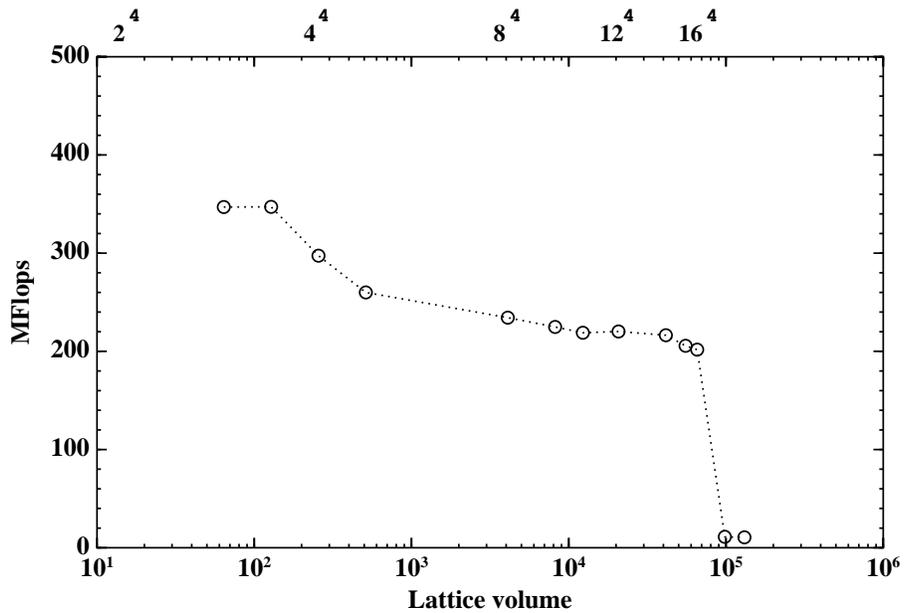}
    \caption{\label{ssor-singlenode}
      The speed of the SSOR preconditioned BiCGStab solver on a single
      node as a function      of the lattice volume.  
      The corresponding lattice dimensions are indicated along the top.
    }
  \end{center} 
\end{figure}

In evaluating the following results it 
will be helpful to know the performance of the solvers on a single node.
Figures \ref{bicgstab-singlenode} and \ref{ssor-singlenode} show how the speed
of the BiCGStab solver and the SSOR preconditioned BiCGStab solver vary with
the number of sites in the lattice. Memory cache and memory bandwidth effects
mean that performance deteriorates as the lattice volume increases.


BiCGStab sustains a speed of 237 Mflops on our $12^4$ test lattice
described in section \ref{testspec}, while
SSOR preconditioned BiCGStab sustains a speed of 220 Mflops.  
The latter solver is 
preferred because it still finds the solution faster due to the smaller number
of iterations.

\subsection{The results of the scaling tests}\label{test results}

We now examine what happens to the performance of our solvers as we
move from a serial to a parallel implementation. There are numerous
factors which might affect the parallel performance;  we make a
preliminary investigation into the effects of data layout and
communication semantics before presenting our final results for the
speedup and parallel efficiency.

\subsubsection{Investigation of data layout}\label{nn-layout-results}

\begin{figure}
  \begin{center} 
    \includegraphics[width=.83\textwidth]{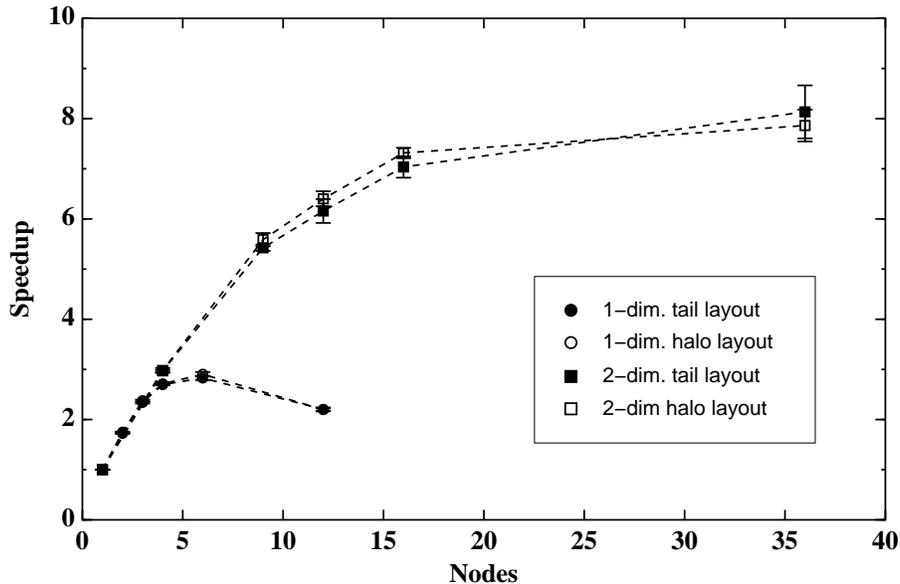}
    \caption{\label{speedup-tail-halo}
      The speedup of the BiCGStab solver using nearest-neighbour blocking
      communications on one- and two-dimensional grids, 
      comparing the data layouts described in section \ref{nn-layout}.
    }
  \end{center} 
\end{figure}

We compare how the two data layouts described in section
\ref{nn-layout} affect the scaling of the data.
It might be expected that the halo layout would have better
spatial locality properties (and therefore use the cache better), 
but the performance, shown in figure
\ref{speedup-tail-halo}, seems to be much the same. This could be
because of the data prefetching written into our assembler kernels.

We choose to use the tail format.  It is easily adapted
to the case of next-to-nearest-neighbour communications (which we need
for the gauge field) by simply adding extra buffer space at the end of
the array. 

\subsubsection{Investigation of communication semantics}\label{blocknonblock}

We can hope to improve performance using non-blocking (asynchronous)
communications (\textit{e.g} \texttt{MPI\_Isend}) rather than
blocking (synchronous) communications (\textit{e.g} \texttt{MPI\_Send}). 

\begin{figure}
  \begin{center} 
    \includegraphics[width=.83\textwidth]{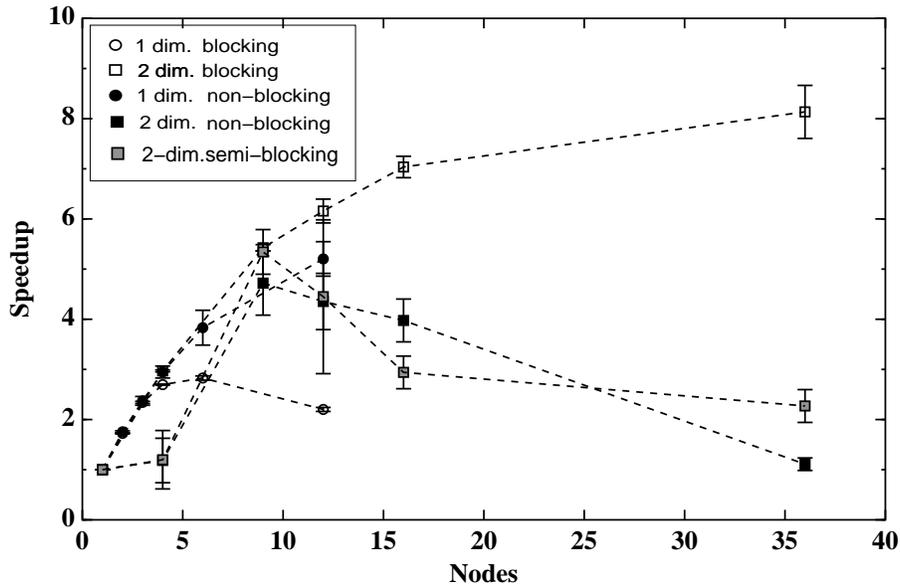}
    \caption{\label{speedup-blocking-nonblocking}
      The speedup of the BiCGStab solver comparing
      nearest-neighbour blocking and non-blocking
      communications.
    }
  \end{center} 
\end{figure}

Figure \ref{speedup-blocking-nonblocking} shows that this improves BiCGStab
performance on a 1-dimensional grid but degrades it on a 2-dimensional grid.
We try a semi-blocking approach where non-blocking communications are used in
direction 0, then a block allows this communication to complete before
non-blocking communications are used in direction 1; this aproach
does not work either.  We have verified that this is also the case for grids
of dimension greater than 2.  Henceforth with BiCGStab we use non-blocking
communications on one-dimensional grids and blocking communications on grids
of more than one dimension.

\begin{figure}
  \begin{center} 
    \includegraphics[width=.83\textwidth]{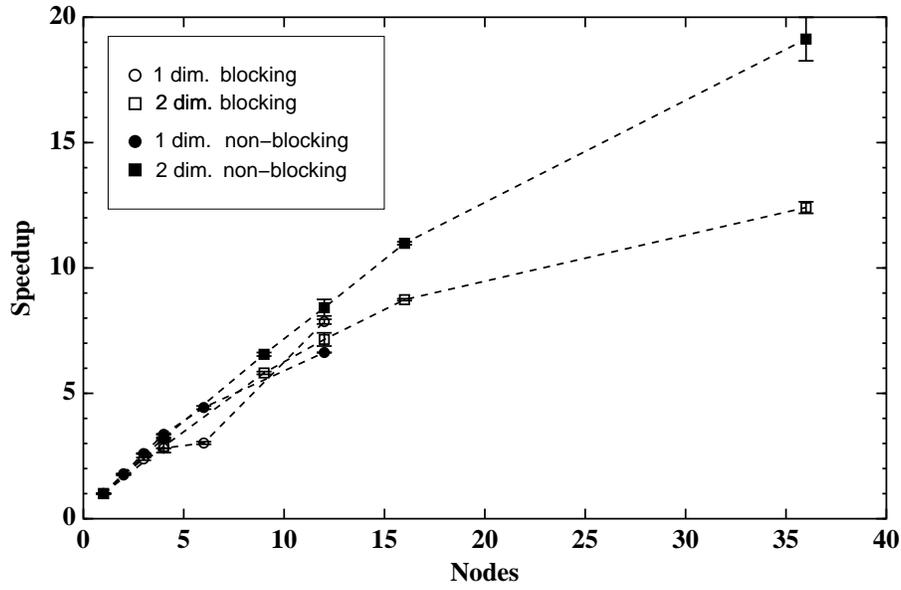}
    \caption{\label{ssor-speedup-blocking-nonblocking}
      The speedup of the SSOR preconditioned BiCGStab solver comparing
      nearest-neighbour blocking and non-blocking
      communications.
    }
  \end{center} 
\end{figure}

Figure \ref{ssor-speedup-blocking-nonblocking} shows that the situation is
different with SSOR preconditioned BiCG\-Stab: non-blocking communications
are faster even on the two-dimensional grid. Henceforth we use non-blocking
communications in connection with SSOR preconditioned BiCGStab.  
We suspect that the communication of BiCGStab, where long messages are sent,
might negatively interfere with caching and compute processes. 
A better separation of communication directions with smaller message
lengths, as is the case for SSOR, might lead less often to a situation
in which data used for compute operations competes for cache lines with
 data used simultaneously for communication 

\subsection{Results for solver speedup and parallel efficiency}

Having determined that the data layout is not very significant and
when it is better to use blocking or non-blocking communications, we
can present the results for the best scaling of our solvers on one- to
four-dimensional grids.

\begin{figure}
  \begin{center} 
    \includegraphics[width=.83\textwidth]{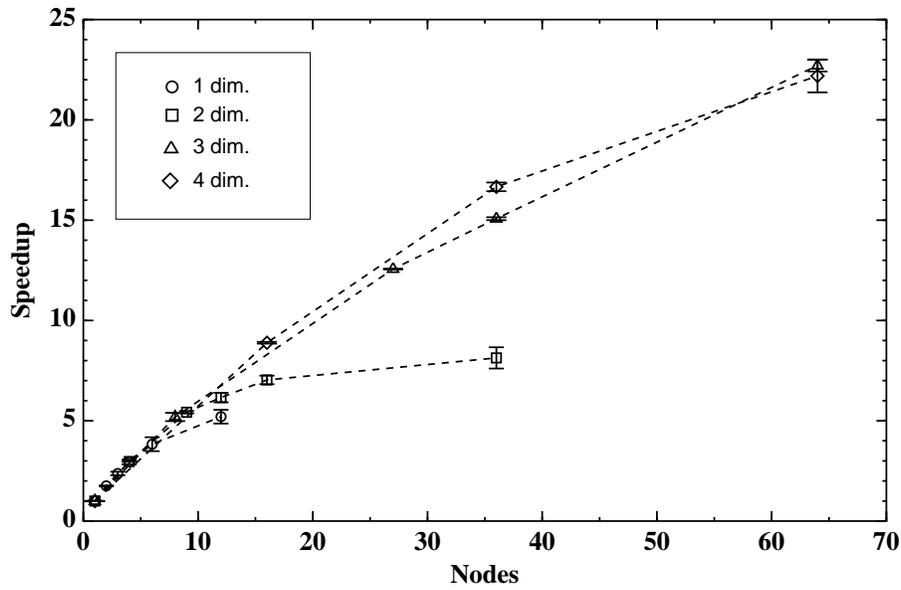}
    \caption{\label{speedup-1234d}
      The speedup of the  BiCGStab solver.
    }
  \end{center} 
\end{figure}
\begin{figure}
  \begin{center} 
    \includegraphics[width=.83\textwidth]{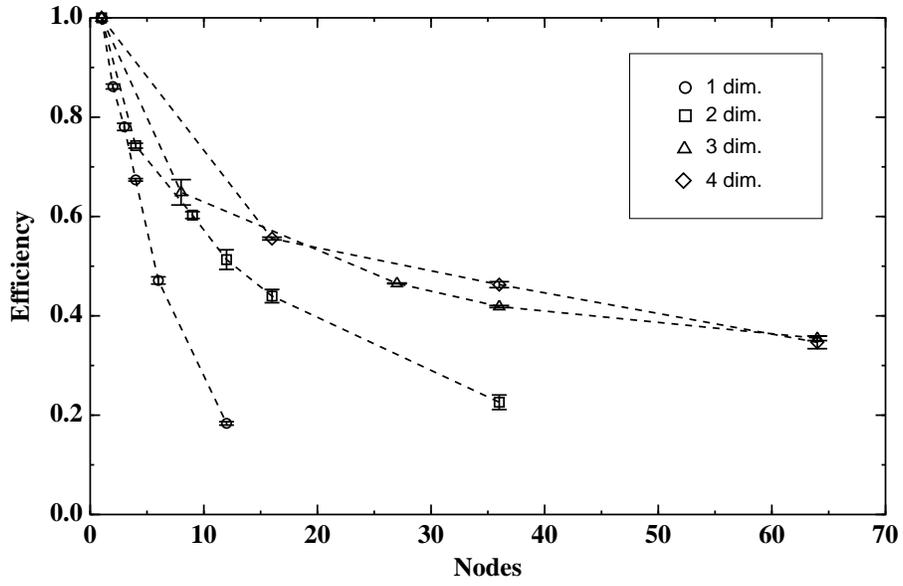}
    \caption{\label{pareff-1234d}
      The parallel efficiency of the BiCGStab solver.
    }
  \end{center} 
\end{figure}

\begin{figure}
  \begin{center} 
    \includegraphics[width=.83\textwidth]{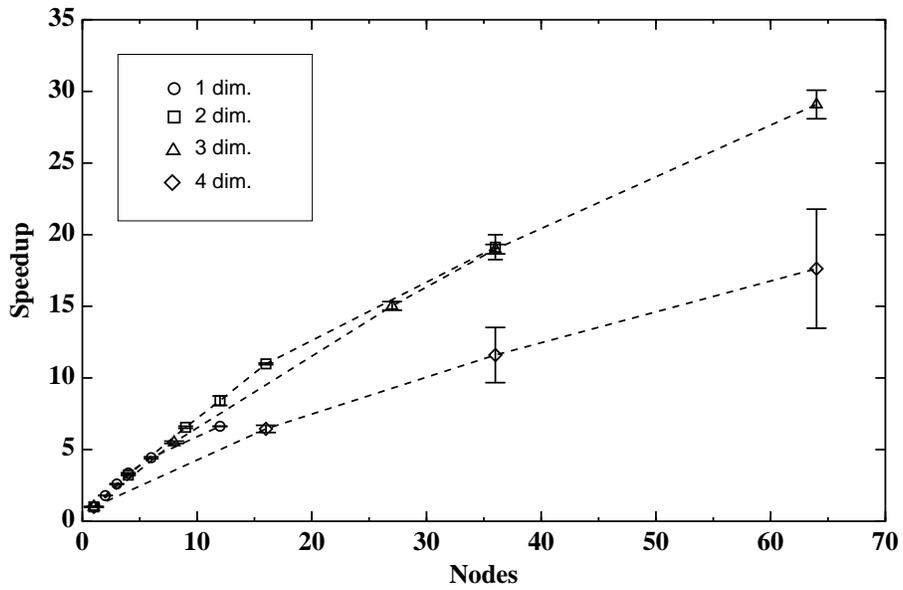}
    \caption{\label{ssor-speedup-1234d}
      The speedup of the SSOR preconditioned BiCGStab solver.
    }
  \end{center} 
\end{figure}
\begin{figure}
  \begin{center} 
    \includegraphics[width=.83\textwidth]{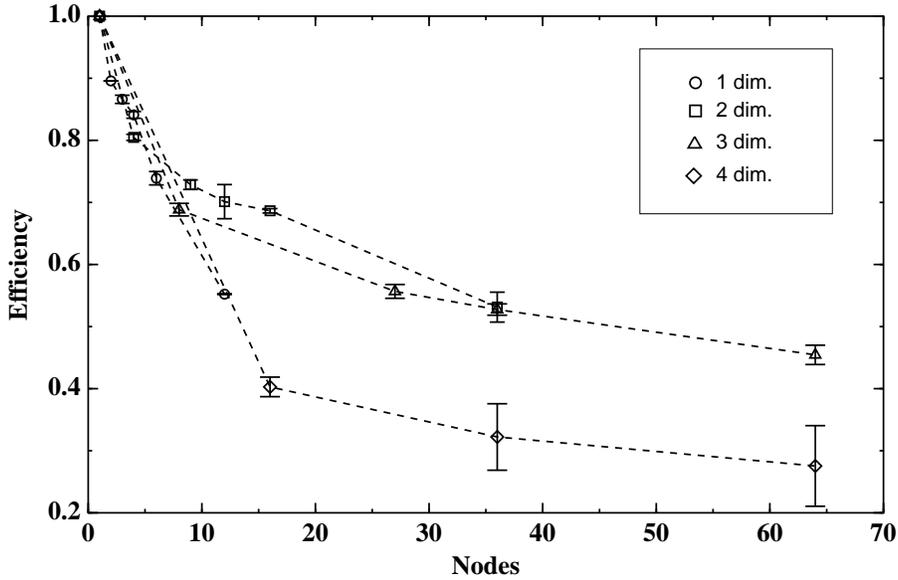}
    \caption{\label{ssor-pareff-1234d}
      The parallel efficiency of the SSOR preconditioned BiCGStab solver.
    }
  \end{center} 
\end{figure}

What we see is that for the BiCGStab solver on this test lattice 
(figures \ref{speedup-1234d} and \ref{pareff-1234d})
the one- and two-dimensional grids become quite inefficient on more
than a few nodes. On the three- and four-dimensional grids the drop
in efficiency on large numbers of nodes is not as dramatic; a
reasonable speedup is achieved due to the more favourable
surface/volume ratio. The speedup on a four-dimensional grid is not
convincingly better than that on a three-dimensional grid, presumably
because the data on the direction 3 boundaries is so finely strided as
to negate the advantage of better surface/volume ratios.

The SSOR results (figures \ref{ssor-speedup-1234d} and
\ref{ssor-pareff-1234d}) show that the speedup is less affected by the
dimensionality of the grid \textit{i.e.} the surface/volume ratio.
However the fact that the data on the direction 3 boundaries are so finely
strided adversely affects the four-dimensional grid performance, since
in this case it means that very many send/receives have to be done,
which is evidently  too inefficient due to latency effects.

A main result of these investigation is that a small $12^4$ lattice still
performs with a scaled speedup of 0.45 on 64 processors. We will comment on
this in section \ref{subsummary}.

\begin{figure}
  \begin{center} 
    \includegraphics[width=.83\textwidth]{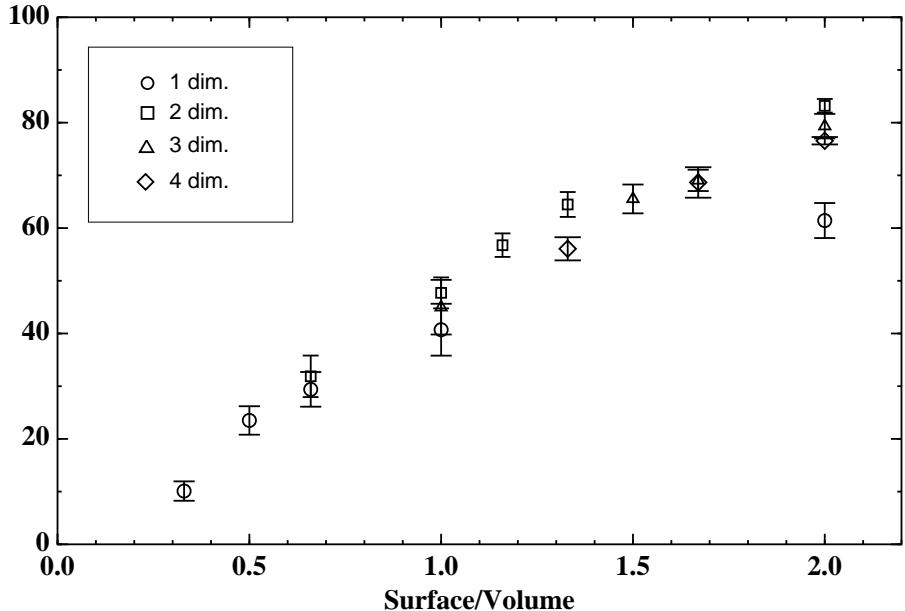}
    \caption{\label{svr-commtime}
      The dependence of the percentage of BiCGStab run-time spent in
      communication on the surface/volume ratio of the local lattice.
    }
  \end{center} 
\end{figure}

\begin{figure}
  \begin{center} 
    \includegraphics[width=.83\textwidth]{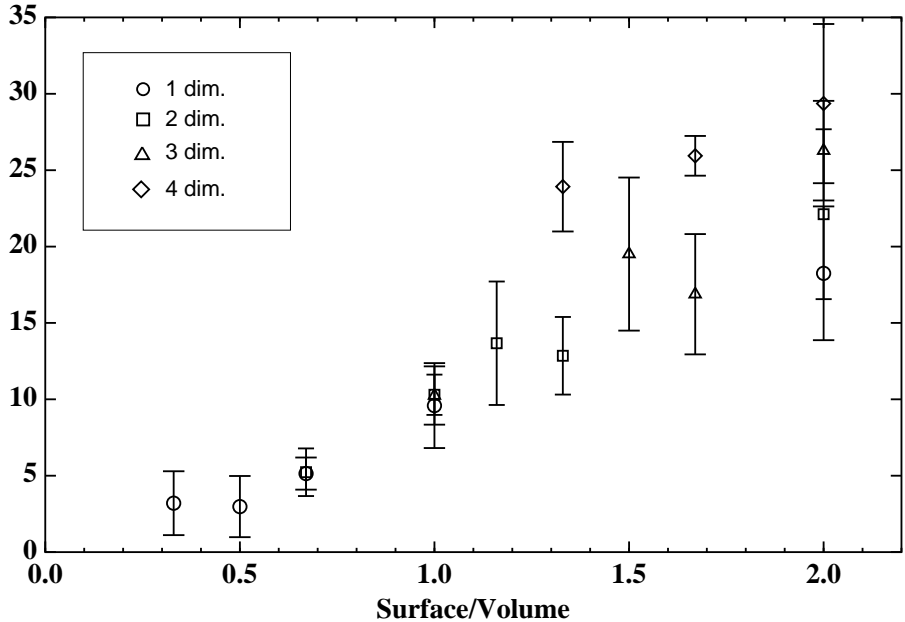}
    \caption{\label{ssor-svr-commtime}
      The dependence of the percentage of SSOR preconditioned 
BiCGStab run-time spent in
      communication on the surface/volume ratio of the local lattice.
    }
  \end{center} 
\end{figure}

To shed further light on these results we plot in figures \ref{svr-commtime}
and \ref{ssor-svr-commtime} the percentage of the total wall-clock execution
time 
required for communication
against the ratio of the communicable surface of the local lattice to the
local lattice volume.  The correlation does not usually depend very strongly
on the dimensionality of the grid; a notable exception is the SSOR
preconditioned solver on the four-dimensional grids.  Note also how the
dependence is weaker when the solver is SSOR preconditioned.


\subsection{Results for next-to-nearest-neighbour communication}\label{nnncomms} 
We examine the overhead introduced by the need to send the extra data
from the boundary corners by implementing next-to-nearest-neighbour
communications and timing exactly the same BiCGStab solve of the same fermion
matrix as before on two-dimensional grids. Any difference is therefore
entirely due to the extra comunication.
Again it was found that for BiCGStab, blocking communications were
fastest, so these tests are performed with blocking communications.

\subsubsection{Investigation of data layout}\label{nnn-layout-results}

\begin{figure}
  \begin{center} 
    \includegraphics[width=.83\textwidth]{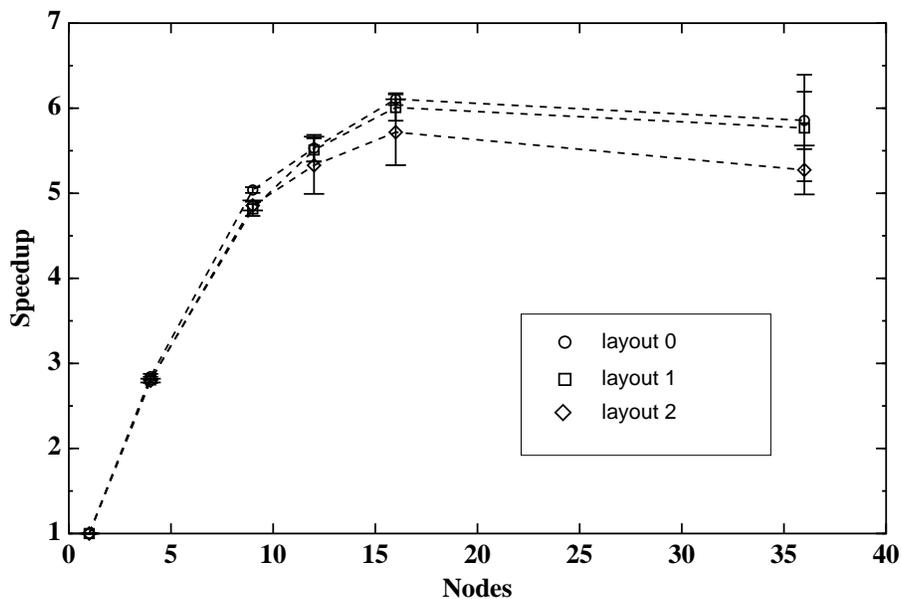}
    \caption{\label{speedup-nnn-layout}
      The speedup of the BiCGStab solver using
      next-to-nearest-neighbour blocking
      communications on  two-dimensional grids, comparing the 
      data layouts described in section \ref{nnn-layout}. 
    }
  \end{center} 
\end{figure}

We  compare in figure \ref{speedup-nnn-layout} the three layouts
 defined in section \ref{nnn-layout}. 
 Because of the seperate
messages required, the third layout is the least scalable,
but the differences only become significant in the
region of poor scaling. Because of this and the fact that for our
applications the next-to-nearest-neighbour communications are
relatively unimportant, we adopt the third layout for the gauge field
in the \gral code, utilising the benefits of a unified data layout
as described in section \ref{nnn-commreq}.

\subsubsection{Results for the speedup}

\begin{figure}
  \begin{center} 
    \includegraphics[width=.83\textwidth]{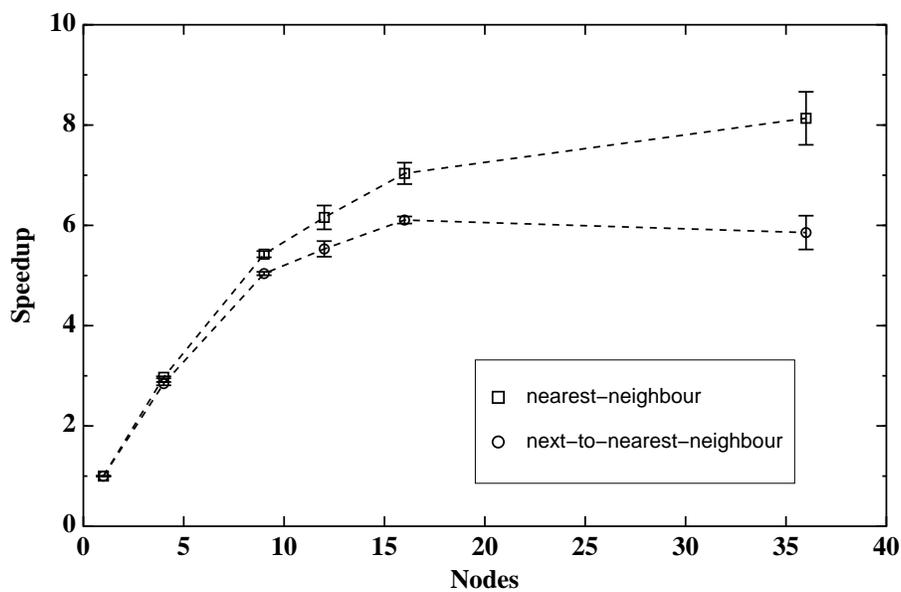}
    \caption{\label{speedup-nnn-vs-nn}
      The speedup of the BiCGStab solver using
      nearest- and next-to-nearest-neighbour blocking
      communications on  two-dimensional grids
    }
  \end{center} 
\end{figure}

In figure \ref{speedup-nnn-vs-nn} we take the most scalable
next-to-nearest-neighbour layout and compare its speedup with the
nearest-neighbour communications on a two-dimension\-al grid, demonstrating
the effect of sending the extra data from the corners of the lattice.

This is perhaps a favourable comparison since
the amount of extra data that must be sent for
next-to-nearest-neighbour communications increases with the
dimensionality $d$ of the node grid as described in section \ref{nnn-commreq}

\section{Summary and  conclusions}\label{summary}

We have presented the results of tests of the scaling 
of the Wilson matrix
solver for a $12^4$ test lattice 
on \alice.
Here we summarize those results and discuss their implications.

\subsection{Summary\label{subsummary}}

On small numbers of nodes the BiCGStab
solver scales without much dependence on the grid decomposition. 
If a minimum effiency of
75\% percent is demanded then figure \ref{pareff-1234d} indicates that no more
than 4 nodes should be used. However it still maintains around 50\% efficiency
on about 16--36 nodes on a three- or four-dimensional grid, corresponding to
local lattices of $6^4$--$4^2\times6^2$ or surface/volume ratios of
1.33--1.67.  Thereafter the efficiency degrades, but not too sharply.

The SSOR preconditioned solver has similar small grid characteristics;
figure \ref{ssor-pareff-1234d} suggests that the 75\% efficiency bound
is reached at around 4--9 nodes. 
The four-dimensional grids rapidly become very slow but an efficiency
of around 50\% is maintained on two and three-dimensional grids of
around 36 nodes, \textit{i.e.} a $4\times4\times3\times12$ local lattice with
surface/volume ratio 1.67 or a $2^2\times12^2$ lattice with surface/volume ratio 2.00.
The performance degrades slowly thereafter; we still  have 45\%
efficiency on 64 nodes with a $3^3\times12$ local lattice.

That the scaling performance need not be seriously affected by the data
layout is a heartening result for code developers; we are free to
organise data in whichever way is convenient. 

A curious feature of our system is the performance of the non-blocking
communications. It appears that if the communications in different
directions are well separated then non-blocking semantics are an
improvement. But they are a hindrance if those communications are done
at the same time, even when separated by
an explicit
barrier as in the semi-blocking case. We have argued above that 
cache effects might be responsible for these differences.

One gratifying result is how well the SSOR preconditioned solver
scales. Its algorithmic performance is excellent but its more
complicated pattern of communication can  in
practice lead to scaling problems due to the communication latency.
The vice of needing many smaller messages rather 
than a few large ones appears to become a virtue, since the separation
of sends allows the faster non-blocking
communications to be used. This demonstrates that a more complicated
algorithm need not be deleterious to efficiency and performance.

\subsection{Discussion}

We see that in general the scaling ability is limited by the
additional time that has to be used for the communications and this is
essentially determined by the surface/volume ratio. Therefore the results
presented here can serve as a guide for planning future runs on
realistic lattices 
of larger global size, bearing in mind that increasing the global size 
introduces additional overheads from the collective communications.

For example, the discussion in the previous section concluded that our
$12^4$ lattice runs with 75\% efficiency on 4 nodes with
BiCGstab. According to table \ref{partab} this corresponds to a
surface/volume ratio of $\frac{2}{3}$. Therefore if we require 75\%
efficiency on any lattice we should aim for grid decomposition which
gives a local surface/volume ratio of about $\frac{2}{3}$. From
equation \eqref{commsurface} we obtain the constraint on the
dimensions of the local lattice in parallel directions 
\be
\sum_{\mu=0}^{d-1}\frac{1}{N_\mu} \simeq \frac{1}{3}.
\ee
This is achieved for a $12^3\times24$ lattice with a grid of
$1\times2\times2\times2$ 
(8 processors)
or possibly $1\times1\times2\times6$ (12 processors).
A $16^3\times32$ lattice needs a $1\times2\times2\times2$ 
(8 processors)
or 
$1\times2\times2\times4$ 
(16 processors)
grid.
A $24^3\times48$ lattice needs a $2\times2\times2\times4$ 
(32 processors)
grid.

With SSOR preconditioning we might be able to use 1.0 as a target
surface/volume ratio, and use grids of $1\times2\times2\times4$ (16),
$1\times2\times4\times4$ (32) and $1\times4\times4\times4$ (64) respectively.
A similar analysis suggests that
a lattice of size $24^4$ can be expected to run with a scaled efficiency
of about 0.45 on 512 processors.

While there must be an optimum grid
decomposition for a given problem and problem size, 
the  complicated nature of the
relationship between latency, bandwidth and message length makes
finding this ``sweet spot'' a difficult optimisation problem.

In general there is a trade-off between speed and efficiency which
must be determined according to the needs of each project and the resources
available; one would also take into account the single-node speed when
deciding the local lattice size.

The \gral project typically requires a number of concurrent 
simulations at different physical parameters which will run for some
considerable length of time. The former requirement will probably mean using a
modest number of nodes on \alice for each run, it being a shared multi-purpose
facility, and the latter requirement will favour running at a reasonable
efficiency. Fortunately these considerations are entirely compatible and we
have shown here that the required performance is available. We also now know
that should a run need to proceed at greater speed, albeit less efficiently,
then this is also feasible.

The target surface/volume ratio will in general change when a different sort
of fermion matrix is used.  Our measurements indicate that a fermion matrix
with next-to-nearest-neighbour communications might introduce an additional
communications overhead of about 10\%.

\subsection{Outlook}

It is reasonable to speculate about what these results mean, or how
they would change, for future cluster computer platforms.
A faster CPU will increase the proportion of time spent in
communications relative to computation, but it should be remembered
that a faster CPU will speed up the communications too, so the overall
effect might not be too different. 

One should be aware that network and communications technology is also
developing: we can look forward to high-bandwidth PCI buses and faster
networks \cite{pcix, infiniband}. In fact much of this technology
exists already, and as cluster computing becomes more widespread
and new network standards (\textit{e.g.} InfiniBand) become available,
current higher-performance technologies (\textit{e.g.} Gigabit System
Network) will become affordable.

A greater network speed helps scalability, but also of importance for the
developer, in that it could change the way application code is designed, are
implementations of communications APIs that make good use of DMA technology to
provide support for simultaneous sends/receives of messages to multiple nodes,
or for overlapping communications with computations.

Of great importance to performance of scientific codes are factors like data
cache sizes and memory bandwidths. If these are greater then it is more
difficult to achieve efficient scaling, but the benefits to the performance on
each node might make decomposition still worthwhile.  Conceivably, there could
then be a situation where a finer grained decomposition actually improves
speedup by moving into a \textit{r\'egime} where all local data fit into
cache.


\section*{Acknowledgements}
ZS acknowledges the financial support provided through the
European Community's Human Potential Programme under contract
HPRN-CT-2000-00145, Hadrons/Lattice QCD. We thank Guido Arnold 
 for support with the ALiCE cluster.
The project has been supported by the Deutsche Forschungsgemeinschaft 
under contract Li701/3-1, {Optimierung von Leistung und Datenorganisation von
  Clusterrechnern der 150 Gflops-Klasse}.



\begin{thebibliography}{8}

\bibitem{top500} \texttt{http://www.top500.org}
\bibitem{rajan} \journal{R.~Gupta}{Parallel Computing}{25}{1999}{1199}
  (\texttt{hep-lat/9905027}) and references therein.
\bibitem{suss} C.~Davies, in \textsl{Heavy Flavour Physics}; Scottish Graduate Textbook Series, Institute of Physics 2002, eds C.~T.~H.~Davies and S.~M.~Playfer
\texttt{hep-lat/0205181}
\bibitem{wilson} K.~G.~Wilson, in \textsl{New Phenomena in Subnuclear
Physics}, ed.~A.~Zichichi (Plenum, New York, 1977)
\bibitem{clover}\journal{B. Scheikholeslami and R. Wohlert}{Nucl.~Phys.}{B 259}{1985}{572}
\bibitem{staggered}\journal{J. Kogut and L. Susskind}{Phys.~Rev.}{D 11}{1975}{395}
\bibitem{morestaggered} \journal{K. Originos \textit{et
    al.}}{Phys.~Rev.}{D 60}{1999}{054503} (\texttt{hep-lat/9903032})
\bibitem{yetmorestaggered}\journal{J-F. Laga\"e and
  D. K. Sinclair}{Phys.~Rev.}{D 59}{1999}{014511}
  (\texttt{hep-lat/9806014})
\bibitem{perfect}\journal{P. Hasenfratz}{Nucl.~Phys. (proc.~suppl.)}{B
  63}{1998}{53} (\texttt{hep-lat/9803027}) and references therein.
\bibitem{gw}\journal{D. B. Kaplan}{Phys.~Lett.}{B 288}{1992}{342}
\bibitem{overlap}\journal{R. Narayanan and
  H. Neuberger}{Nucl.~Phys.}{B 433}{1995}{305}
\bibitem{gral}\journal{B.~Orth \textit{et
    al.}}{Nucl.~Phys.~(proc.~suppl)}{106}{2002}{289} (\texttt{hep-lat/0110158})
\bibitem{akmt} 
\texttt{http://www.theorie.physik.uni-wuppertal.de/Computerlabor/Alice/akmt.phtml}
\bibitem{bicgstab}
\journal{H. A. Van der Vorst}{SIAM J.~Sci.~Stat.~Comp.}{13}{1992}{631}
\bibitem{morebicgstab}\journal{A. Frommer \textit{et
    al.}}{Int.~J.~Mod.~Phys.}{C 5}{1994}{1073}
\bibitem{llssor}\journal{S. Fischer \textit{et
    al.}}{Comp.~Phys.~Comm.}{98}{1996}{20}
\bibitem{morellssor}\journal{Th.~Lippert}{Parallel
  Computing}{25}{1999}{1357}
\bibitem{alice} \journal{N. Eicker \textit{et
    al.}}{Nucl.~Phys.~(proc.~suppl.)}{B 83}{2000}{798} (\texttt{hep-lat/9909146})\\
\texttt{http://www.theorie.physik.uni-wuppertal.de/Computerlabor/ALiCE.phtml}
\bibitem{suse} \texttt{http://www.suse.de}
\bibitem{parastation} \texttt{http://www.par-tec.de}
\bibitem{parallelfilesystem}
Th.~D\"ussel \textit{et al.}; \texttt{cs.DC/0303016}. To be published in
\textsl{Parallel Computing.}
\bibitem{pcix}\texttt{http://www.pcisig.com/specifications/pcix\_20}
\bibitem{infiniband}\texttt{http://www.infinibandta.org}
\end{thebibliography}
\end{document}